\documentclass[twoside,a4paper,fleqn]{mysvjour2}                    
\smartqed  
\usepackage{graphicx}\usepackage{amsmath,mathrsfs}
\usepackage{epsfig}\usepackage{psfrag}
\usepackage{subfigure}
\usepackage{tabularx}\usepackage{rotating}
\def\figurename{Fig.}

\addtocounter{footnote}{0}

 \usepackage{mathptmx}      
\usepackage{latexsym}
 \journalname{Celestial Mechanics and Dynamical Astronomy (2007) 97:249-265}
\begin{document}
\markboth{Z.~Jiang, L.~Ossipkov}{Anisotropic distribution functions for spherical galaxies}

\title{Anisotropic distribution functions for spherical galaxies
}

\titlerunning{Anisotropic distribution functions for spherical galaxies}        

\author{Zhenglu Jiang         \and
        Leonid Ossipkov  
}

\authorrunning{Z.~Jiang \& L.~Ossipkov} 

\institute{Z.~Jiang \at
              Department of Mathematics, Zhongshan University, 
              Guangzhou 510275, China \\
              \email{mcsjzl@mail.sysu.edu.cn}           
           \and
           L.~Ossipkov \at
              Saint~Petersburg State University, Staryj Peterhof, Saint~Petersburg 198504, Russia \\
          \email{leo@dyna.astro.spbu.ru}
}

\date{Received: 30 October 2006 / Accepted: 23 December 2006 / Published online: 9 February 2007}

\maketitle

\begin{abstract}
A method is presented for finding anisotropic distribution functions
for stellar systems with known, spherically symmetric, densities,
which depends only on the two classical integrals of the energy and the 
magnitude of the angular momentum. 
It requires the density to be expressed as  
  a sum of products of functions of the potential and 
  of the radial coordinate. 
The solution corresponding to this type of density  
is in turn a sum of products of functions of the energy and of the 
magnitude of the angular momentum. 
The products of the density and its radial and transverse velocity dispersions   
can be also expressed as  
  a sum of products of functions of the potential and 
 of the radial coordinate.   
Several examples are given, including some of new 
anisotropic distribution functions. 
This device can be extended further to the related problem of 
finding two-integral distribution functions for axisymmetric galaxies.
\end{abstract}
 
\section{Introduction}
\label{intro}
There is a long history of studying the structure of galaxies by  
construction of self-consistent distribution functions 
for a stellar system with a known gravitational potential.  The potential of 
the system determines the self-consistent 
mass density $\rho$ of the system via Poisson's 
equation generated by the well-known Newtonian gravitational law, and also 
the structure of the stellar orbits according to Newton's 
equations of motion. The system is constructed from 
the building blocks of the orbits 
that can lie within the potential. 
This is called the ``from $\rho$ to $f$'' approach for finding a 
self-consistent distribution function $f$ (Binney and Tremaine 1987, hereafter BT).
The distribution function (hereafter DF) of the system 
represents how stars are distributed 
in the phase space of the system   
and the integration of the DF over velocity space yields the density.  
Therefore the problem of finding the DF is that of solving an 
integral equation; 
in the spherical case in particular this equation is 
of the first kind, in which the unknown DF $f$ 
only occurs inside the integral. 

Some outstanding astronomers have contributed to 
this mathematical problem. 
Eddington (1916) showed that it can be solved for an isotropic 
DF that depends only on the energy in the spherical case, 
whose velocity dispersions in the radial and tangential directions must be equal,
by first expressing the density as a function only of the potential, 
and then solving the Abel integral equation. 
It is also well known that the isotropic DF is unique for the potential of 
any spherical system. 

Fricke's (1952) expansion method can be obviously 
generalized to the spherical case (e.g., Camm 1952; Bouvier 1962, 1963; 
Ossipkov 1979a; Kent and Gunn 1982; 
Dejonghe 1986, 1987; Dejonghe and Merritt 1988; Cuddeford 1991; Louis 1993),
and similarly yields the result that 
DFs which are products of the two powers 
of the energy and the square of the angular momentum 
correspond to densities which are proportional to products of the potential 
and the spherical radial coordinate for spherical systems. 
Hence the DF for the system can be obtained 
by first expressing the density as a function of 
the potential and the spherical radial coordinate, 
and then expanding as a power series (Qian and Hunter 1995). 
Moreover, there may be an infinity of anisotropic DFs corresponding to 
any given mass density  in spherical stellar systems (Dejonghe 1987).  

A class of typical anisotropic DFs that depend on the energy and 
the magnitude of the angular momentum,  whose velocity dispersions 
are anisotropic (but ellipsoidal), was independently found by Ossipkov (1979) 
and Merritt (1985), for a spherical density distribution. 
Such anisotropic DFs  of Ossipkov-Merritt type
are in fact  analogues of Eddington's formula, mentioned above.
In a number of papers, different integral transformation techniques can be used to 
obtain the solution of the spherical (or axisymmetric) problem 
(e.g. Lynden-Bell 1962; Hunter 1975; Kalnajs 1976; Dejonghe 1986; 
 Qian and Hunter 1995) but 
there is the same difficulty of requiring not only 
the validity of these transformations of the density 
but also the complex analyticity 
of a density-related integral kernel to complex arguments. 
The contour integral method of Hunter and Qian (1993) can be also 
used to find anisotropic DFs for spherical systems (Qian and Hunter 1995) 
but it is valid for densities that are analytic, and whose  
singularities satisfy some conditions.   

It is worth mentioning that Kuzmin and Veltmann (1967a, 1973) and 
Veltmann (1961, 1965, 1979, 1981) developed some more 
general classes when DF is a product of an unknown function over one argument  
on a known function of another.

The fundamental integral equations of the problem are given 
in {Sect.} \ref{finteq}, and some new formulae 
for finding other anisotropic DFs for stellar systems
with known spherically symmetric densities are presented 
in {Sect.} \ref{anisdf}.
These depend only on the two integrals of the energy and the 
 magnitude of the angular momentum.   
This work constitutes the core of this paper.  
These formulae in fact come from 
an combination of the ideas of 
 Eddington and Fricke (see above).  Of course, 
they can be also regarded as simply
an extension of Eddington's formula. 
A type of anisotropic DF which is  
a sum of products of functions only of the energy  
and powers of the magnitude of the angular momentum 
is derived in {Sect.} \ref{anisdf1} and  
another, which is a sum of products 
of functions only of a special variable  
and powers of the magnitude of the angular momentum,
in {Sect.} \ref{anisdf2}. 
More general formulae are given in the last part of {Sect.} \ref{anisdf}.  
Various formulae of the velocity dispersions for such models of these DFs 
are also  shown in all the three parts of {Sect.} \ref{anisdf}. 
Several examples are given in {Sect.} \ref{appli}, including some of new 
anisotropic DFs.  Different anisotropic DFs for the Plummer model
 are first given in {Sect.} \ref{apm}. 
Then anisotropic DFs of the H\'{e}non isochrone model 
appear in {Sect.} \ref{ahm} 
and those of the $\gamma$-model are described in {Sect.} \ref{agm}.  
Section {\ref{con} is a summary and conclusion.

\section{The fundamental integral equations}
\label{finteq}
For convenience of calculation, as in BT it is usual to introduce 
the relative potential $\psi$ and the relative energy $\varepsilon$ 
of a star in a stellar system, defined by 
$\psi=-\Phi+\Phi_0$ and $\varepsilon=-E+\Phi_0,$ where 
$\Phi$ and $E$ are, respectively, the potential 
and the energy of a star,  and $\Phi_0$ is a constant 
generally chosen to be such that 
there are only stars of $\varepsilon>0$ in 
the system modelled by DFs. 
It is well known that the relative energy $\varepsilon$  
and the three components of the angular momentum vector ${\bf L}$ 
are four isolating integrals for any orbit in a spherical potential. 
By the Jeans theorem, it follows that the  
 DF of a steady-state spherical stellar system 
can be regarded as a non-negative function of these integrals, 
denoted by $f=f(\varepsilon,{\bf L}).$ 
If the system is spherically symmetric in all its properties, 
$f$ is independent of the direction of ${\bf L}$ and depends only on
its magnitude $L$ (e.g., Shiveshwarkar 1936; Ogorodnikov 1965), that is, $f$ can be expressed 
as a non-negative function of the relative energy $\varepsilon$  
and the absolute value $L$ of the angular momentum vector ${\bf L},$ 
denoted by $f=f(\varepsilon, L).$ 
If the stellar system itself provides the relative potential $\psi=\psi({\bf r}),$  
the mass density $\rho=\rho({\bf r})$ can be obtained from Poisson's equation,
and it is known that its distribution function $f=f({\bf r},{\bf v})$ satisfies 
\begin{equation}
-\nabla^2\psi =4\pi G \rho=4\pi G\int fd^3{\bf v}, \label{int0}
\end{equation}
where ${\bf r}$ is a position vector, ${\bf v}$ is a velocity vector, 
$G$ is the gravitational constant. Let $r$ be the modulus of the position vector ${\bf r}.$ 
For a spherically symmetric system, both the relative potential $\psi=\psi({\bf r})$ 
and the mass density $\rho=\rho({\bf r})$ can be, respectively, regarded as 
two functions of $r,$ that is, $\psi=\psi(r)$ and $\rho=\rho(r),$ and the distribution function $f=f({\bf r},{\bf v})$ 
can be expressed as a function of the relative energy $\varepsilon$ and the momentum magnitude $L,$ thus 
equation (\ref{int0}) can be rewritten as 
\begin{equation}
-\frac{1}{r^2}\frac{d}{dr}\left(r^2\frac{d\psi}{dr}\right)
 =4\pi G \rho=4\pi G\int f(\varepsilon,L)d^3{\bf v}. \label{int1}
\end{equation}
Once $f(\varepsilon,L)$ is known, $\rho$ can be easily calculated 
by integration and $\psi$ by solving Poisson's equation for the spherical system. 
 The inverse problem that  is now investigated is 
how to derive the DF from the density $\rho$ for any spherical system.  
Different classes of anisotropic DFs will be below shown,  
which are derived from spherical density profiles for galaxies,  
by combining some functions only of $\varepsilon$ 
(or $Q\equiv\varepsilon-L^2/(2r_a^2)$) 
with some functions of the form $L^{2n\beta_n}$ 
where $r_a$ is a scaling radius, $n$ is an integer greater than $-2$ 
and $\beta_n$ is a constant such that $n\beta_n>-1.$ 

\section{Anisotropic DFs} 
\label{anisdf}
In this section various formulae for anisotropic DFs are obtained 
from spherical density profiles of different forms 
and their radial and transverse velocity dispersions are expressed 
in a simple form. 
We will consider only infinite-size models, so $\varepsilon=0$ will be an energy of escape from a system. 

\subsection{DFs of the form $\sum\limits_{n=0}^{m}L^{2n}h_n(\varepsilon)$} 
\label{anisdf1}
The integral on the right side of  (\ref{int1}) is first expressed
in velocity space. 
Let ${\bf v}=(v_r,v_\varphi,v_\theta)$ be the velocity 
in the spherical coordinates $(r,\varphi,\theta)$ of space 
and $v_T=\sqrt{v_\varphi^2+v_\theta^2}$ be the transverse velocity. 
Then $L=rv_T$ and, by $\varepsilon=\psi-(v_r^2+v_T^2)/2,$  (\ref{int1}) can be expressed as
\begin{equation}
 \rho(r)=4\pi\int_0^\psi \left[\int_0^{\sqrt{2\varepsilon}}\frac{f(\psi-\varepsilon,rv_T)v_T}
{\sqrt{2\varepsilon-v_T^2}}dv_T\right]d\varepsilon  
 \label{int2}
\end{equation}
for the spherical density $\rho=\rho(r)$ since 
the system has only stars with $\varepsilon>0,$ or say, $f(\varepsilon,L)=0$ for $\varepsilon\leq 0.$ 
This means that the mass density $\rho(r)$ can be regarded as a function 
depending on the radial coordinate $r$ and the relative potential $\psi.$ 
Let  $\rho(r)$ be below denoted by $\rho(\psi,r).$ 
Assume that $f(\varepsilon,L)=\sum\limits_{n=0}^{m}L^{2n}h_n(\varepsilon)$ 
for  $\varepsilon>0$ and $f(\varepsilon,L)=0$ for $\varepsilon\leq 0,$ and that 
the density $\rho(\psi,r)$ has the form of 
$\rho(\psi,r)=\sum\limits_{n=0}^{m}\tilde{\rho}_n(\psi)r^{2n}$ 
corresponding to the relative potential  $\psi.$  Notice that such representation of the density is not unique. 
Similar models were discussed earlier by Bouvier (1962, 1963). 
By (\ref{int2}), it follows that 
\begin{equation}
 \sum\limits_{n=0}^{m}\tilde{\rho}_n(\psi)r^{2n}=\sum\limits_{n=0}^{m}\frac{(2\pi)^{3/2}2^nn!r^{2n}} 
{\Gamma(n+3/2)}\int_0^\psi h_n(\varepsilon)(\psi-\varepsilon)^{n+1/2}d\varepsilon.  
 \label{int3}
\end{equation}
Equation (\ref{int3}) holds if $h_n(\varepsilon)$ ($n=0,1,2,\dots,m$) satisfy 
the conditions
\begin{equation}
 \tilde{\rho}_n(\psi)=(2\pi)^{3/2}\frac{2^nn!} 
{\Gamma(n+3/2)}\int_0^\psi h_n(\varepsilon)(\psi-\varepsilon)^{n+1/2}d\varepsilon.   
 \label{int4}
\end{equation}
By taking the $(n+1)$st derivative of  both sides of equation (\ref{int4}) with respect to $\psi$ 
and using Abel's integral equation, 
it is found that equation (\ref{int4}) can be solved by 
\begin{equation}
 h_n(\varepsilon)=\frac{1}{\sqrt{8}\pi^22^nn!}\left[\int_0^\varepsilon
\frac{d^{n+2}\tilde{\rho}_n(\psi)}{d\psi^{n+2}}\frac{d\psi}{\sqrt{\varepsilon-\psi}}
+\frac{1}{\sqrt{\varepsilon}}\left(\frac{d^{n+1}\tilde{\rho}_n(\psi)}{d\psi^{n+1}}\right)_{\psi=0}\right]   
 \label{solution}
\end{equation}
for $n=0,1,2,\dots,m.$   
Hence it can be easily found from (\ref{solution}) that 
\begin{equation}
f(\varepsilon,L)=\frac{1}{\sqrt{8}\pi^2}
\sum\limits_{n=0}^{m}\frac{L^{2n}}{2^nn!}\left[\int_0^\varepsilon
\frac{d^{n+2}\tilde{\rho}_n(\psi)}{d\psi^{n+2}}\frac{d\psi}{\sqrt{\varepsilon-\psi}}
+\frac{1}{\sqrt{\varepsilon}}\left(\frac{d^{n+1}\tilde{\rho}_n(\psi)}{d\psi^{n+1}}\right)_{\psi=0}\right]
\label{df}
\end{equation}
for $\varepsilon>0,$ 
which is a class of anisotropic DFs generated from the spherical density 
of the form $\rho(\psi,r)=\sum\limits_{n=0}^{m}\tilde{\rho}_n(\psi)r^{2n}.$
Furthermore, if $(d^j\tilde{\rho}_n(\psi)/d\psi^j)_{\psi=0}=0$ for 
$j=0,1,\dots,m+1,$ then, for $\varepsilon>0,$ (\ref{df}) can be expressed as 
\begin{equation}
f(\varepsilon,L)=\frac{1}{\sqrt{8}\pi^2}
\sum\limits_{n=0}^{m}\frac{L^{2n}}{2^nn!}\frac{d^{n+2}}{d\varepsilon^{n+2}}\int_0^\varepsilon
\frac{\tilde{\rho}_n(\psi)d\psi}{\sqrt{\varepsilon-\psi}}.
\label{df0}
\end{equation} 

 By (\ref{df}), the velocity dispersions $\sigma_r^2(\psi,r)$ and $\sigma_T^2(\psi,r)$ 
can be also found to be of the following forms 
\begin{equation}
\sigma_r^2(\psi,r)=\frac{1}{\rho(\psi,r)}\sum\limits_{n=0}^{m}r^{2n}\int_0^\psi \tilde{\rho}_n(\psi^\prime)d\psi^\prime
\label{vdr}
\end{equation} 
and 
\begin{equation}
\sigma_T^2(\psi,r)=\frac{1}{\rho(\psi,r)}\sum\limits_{n=0}^{m}(n+1)r^{2n}\int_0^\psi \tilde{\rho}_n(\psi^\prime)d\psi^\prime
\label{vdvt}
\end{equation} 
for any DF derived from the spherical density 
of the form $\rho(\psi,r)=\sum\limits_{n=0}^{m}\tilde{\rho}_n(\psi)r^{2n}.$ It can be also known that 
these dispersions (\ref{vdr}) and (\ref{vdvt}) can be obtained directly
according to the following velocity dispersion formulae (Dejonghe 1986, 1987)  
\begin{equation}
\sigma_r^2(\psi,r)=\frac{1}{\rho(\psi,r)}\int_0^\psi \rho(\psi^\prime,r)d\psi^\prime
\label{vdrd}
\end{equation} 
and 
\begin{equation}
\sigma_\varphi^2(\psi,r)=\sigma_\theta^2(\psi,r)=\frac{1}{\rho(\psi,r)}
\int_0^\psi \frac{\partial [r^2\rho(\psi^\prime,r)]}{\partial r^2}d\psi^\prime. 
\label{vdrd}
\end{equation} 
\subsection{DFs of the form $\sum\limits_{n=0}^{m}L^{2n}g_n(Q)$}
\label{anisdf2}
A more general expression for the integral in the right side of (\ref{int1})
 can also be derived. 
To do this, first put $Q=\psi-(v_r^2+v_T^2)/2-r^2v_T^2/(2r_a^2),$ where $v_r$ and $v_T$ are the same 
as in {Sect.} \ref{anisdf1} and $r_a$ is a scaling radius. 
Obviously, $Q=\varepsilon-L^2/(2r_a^2)$ and $Q\to \varepsilon$ as $r_a \to \infty.$ 
Assume that the DF is of the form $f=f(Q,L),$  
and that the system has only stars with $Q>0,$
or equivalently, $f=0$ for $Q\leq 0.$ 
Then it is easy to see that 
\begin{equation}
 \rho(r)=4\pi\int_0^\psi \left[\int_0^{\sqrt{2Q/(1+r^2/r_a^2)}}\frac{f(\psi-Q,rv_T)v_T}
{\sqrt{2Q-(1+r^2/r_a^2)v_T^2}}dv_T\right]dQ.  
 \label{intq2}
\end{equation}
This shows that the density $\rho(r)$ can be also regarded as a function of 
the radial coordinate $r$ and the relative potential $\psi.$ 
Suppose that $f(Q,L)=\sum\limits_{n=0}^{m}L^{2n}g_n(Q)$ for $Q>0,$ 
and that the density $\rho(r)$ can be expressed as 
$\rho(\psi,r)=\sum\limits_{n=0}^{m}\hat{\rho}_n(\psi)r^{2n}/(1+r^2/r_a^2)^{n+1}.$ 
Then similarly by (\ref{intq2}) it follows that 
\begin{equation}
 \sum\limits_{n=0}^{m}\frac{\hat{\rho}_n(\psi)r^{2n}}{(1+r^2/r_a^2)^{n+1}}=
\sum\limits_{n=0}^{m}\frac{(2\pi)^{3/2}2^nn!r^{2n}} 
{\Gamma(n+3/2)(1+r^2/r_a^2)^{n+1}}\int_0^\psi g_n(Q)(\psi-Q)^{n+1/2}dQ.  
 \label{intq3}
\end{equation}
Equation (\ref{intq3}) holds if  $g_n(Q)$ ($n=0,1,2,\dots,m$) satisfy 
\begin{equation}
 \hat{\rho}_n(\psi)=\frac{(2\pi)^{3/2}2^nn!} 
{\Gamma(n+3/2)}\int_0^\psi g_n(Q)(\psi-Q)^{n+1/2}dQ.   
 \label{intq4}
\end{equation}
By taking the $(n+1)$st derivative of both sides of equation (\ref{intq4}) with respect to $\psi,$ 
and using Abel's integral equation,
the solution of equation (\ref{intq4}) is given by 
\begin{equation}
g_n(Q)=\frac{1}{\sqrt{8}\pi^22^nn!}\left[\int_0^Q
\frac{d^{n+2}\hat{\rho}_n(\psi)}{d\psi^{n+2}}\frac{d\psi}{\sqrt{Q-\psi}}
+\frac{1}{\sqrt{Q}}\left(\frac{d^{n+1}\hat{\rho}_n(\psi)}{d\psi^{n+1}}\right)_{\psi=0}\right]   
 \label{solutionq}
\end{equation}
for $n=0,1,2,\dots,m.$ Then it can be easily shown from (\ref{solutionq}) that 
\begin{equation}
f(Q,L)=\frac{1}{\sqrt{8}\pi^2}
\sum\limits_{n=0}^{m}\frac{L^{2n}}{2^nn!}\left[\int_0^Q
\frac{d^{n+2}\hat{\rho}_n(\psi)}{d\psi^{n+2}}\frac{d\psi}{\sqrt{Q-\psi}}
+\frac{1}{\sqrt{Q}}\left(\frac{d^{n+1}\hat{\rho}_n(\psi)}{d\psi^{n+1}}\right)_{\psi=0}\right]
\label{dfq}
\end{equation}
for $Q>0,$ 
which is another class of anisotropic DFs generated from the spherical density
of the form of $\rho(\psi,r)=\sum\limits_{n=0}^{m}\hat{\rho}_n(\psi)r^{2n}/(1+r^2/r_a^2)^{n+1}.$
Furthermore, if $(d^j\hat{\rho}_n(\psi)/d\psi^j)_{\psi=0}=0$ for 
$j=0,1,2,\dots,m+1,$ then for $Q>0,$ (\ref{dfq}) can be rewritten as 
\begin{equation}
f(Q,L)=\frac{1}{\sqrt{8}\pi^2}
\sum\limits_{n=0}^{m}\frac{L^{2n}}{2^nn!}\frac{d^{n+2}}{dQ^{n+2}}\int_0^Q
\frac{\hat{\rho}_n(\psi)d\psi}{\sqrt{Q-\psi}}.
\label{dfq0}
\end{equation}

Of course,  (\ref{dfq}) and (\ref{dfq0})  coincide with (\ref{df}) and (\ref{df0}), respectively.   
In other words, (\ref{df}) and (\ref{df0})  are, respectively, limits of (\ref{dfq}) and (\ref{dfq0}) when $r_a\to \infty.$ 

Similar to those in {Sect.} \ref{anisdf1}, the velocity dispersions $\sigma_r^2(\psi,r)$ and $\sigma_T^2(\psi,r)$ can be also found to be of the following forms 
\begin{equation}
\sigma_r^2(\psi,r)=\frac{1}{\rho(\psi,r)}\sum\limits_{n=0}^{m}
\frac{r^{2n}}{(1+r^2/r_a^2)^{n+1}}\int_0^\psi \hat{\rho}_n(\psi^\prime)d\psi^\prime
\label{vdrq}
\end{equation} 
and 
\begin{equation}
\sigma_T^2(\psi,r)=\frac{1}{\rho(\psi,r)}\sum\limits_{n=0}^{m}(n+1)\frac{r^{2n}}{(1+r^2/r_a^2)^{n+2}}\int_0^\psi \hat{\rho}_n(\psi^\prime)d\psi^\prime
\label{vdvtq}
\end{equation} 
for  any DF derived from the spherical density 
of the form $\rho(\psi,r)=\sum\limits_{n=0}^{m}\hat{\rho}_n(\psi)r^{2n}/(1+r^2/r_a^2)^{n+1}.$  

It is worth mentioning that if the sum contains only one term then such models have been studied by Cuddeford (1991). 
Therefore the above DFs can be indeed regarded as a simple generalization of the models given by Cuddeford.  
However, (\ref{dfq}) and (\ref{dfq0}) are new formulae for our finding anisotropic DFs. 

\subsection{Miscellaneous DFs}
\label{anisdf3}
One can also obtain more general formulae than (\ref{df}) and (\ref{dfq}).  It can be shown that 
\begin{equation}
f(\varepsilon,L)=
\sum\limits_{n=0}^{m}B_nL^{2n\beta_n}
\left[\int_0^\varepsilon\frac{d^{a_n+1}\tilde{\rho}_n(\psi)}
{d\psi^{a_n+1}}\frac{d\psi}{(\varepsilon-\psi)^{\alpha_n}}
+\frac{1}{\varepsilon^{\alpha_n}}\left(\frac{d^{a_n}\tilde{\rho}_n(\psi)}
{d\psi^{a_n}}\right)_{\psi=0}\right]
\label{dfg}
\end{equation}
for $\varepsilon>0$ is an anisotropic DF for the spherical density of the form 
$\rho(\psi,r)=\sum\limits_{n=0}^{m}\tilde{\rho}_n(\psi)r^{2n\beta_n}$, 
 with $n\beta_n>-1.$ Here
$B_n=[(2\pi)^{3/2}2^{n\beta_n}\Gamma(n\beta_n+1)\Gamma(1-\alpha_n)]^{-1},$ 
$\alpha_n=n\beta_n-a_n+3/2$ and $a_n$ is 
a non-negative integer such that $0\leq\alpha_n<1$ for $n=0,1,\cdots,m.$ 

Similarly, it can be readily shown that a further anisotropic DF is given by 
\begin{equation}
f(Q,L)=
\sum\limits_{n=0}^{m}B_nL^{2n\beta_n}
\left[\int_0^Q\frac{d^{a_n+1}\hat{\rho}_n(\psi)}{d\psi^{a_n+1}}\frac{d\psi}{(Q-\psi)^{\alpha_n}}
+\frac{1}{Q^{\alpha_n}}\left(\frac{d^{a_n}\hat{\rho}_n(\psi)}{d\psi^{a_n}}\right)_{\psi=0}\right]
\label{dfqg}
\end{equation}
for $Q>0,$ corresponding to the spherical density  
$\rho(\psi,r)=\sum\limits_{n=0}^{m}\hat{\rho}_n(\psi)r^{2n\beta_n}/(1+r^2/r_a^2)^{n\beta_n+1},$ 
 where $Q$ is given in {Sect.} \ref{anisdf2}, and $B_n,a_n,\alpha_n$ and $\beta_n$ 
are the same as in (\ref{dfg}). 

Put $\hat{Q}=\max(Q,0).$ Then, it can be furthermore shown that the DFs of the form 
\begin{eqnarray}
f(\varepsilon,Q,L)=&
\sum\limits_{n=0}^{m}B_{1n}L^{2n\beta_{1n}}
\left[\int_0^\varepsilon\frac{d^{a_n+1}\tilde{\rho}_n(\psi)}
{d\psi^{a_{1n}+1}}\frac{d\psi}{(\varepsilon-\psi)^{\alpha_{1n}}}
+\frac{1}{\varepsilon^{\alpha_{1n}}}\left(\frac{d^{a_{1n}}\tilde{\rho}_n(\psi)}
{d\psi^{a_{1n}}}\right)_{\psi=0}\right] \hspace*{2.5cm} \nonumber \\
& + \sum\limits_{n=0}^{m}B_{2n}L^{2n\beta_{2n}}
\left[\int_0^{\hat{Q}}\frac{d^{a_{2n}+1}\hat{\rho}_n(\psi)}{d\psi^{a_{2n}+1}}\frac{d\psi}{(\hat{Q}-\psi)^{\alpha_{2n}}}
+\frac{1}{\hat{Q}^{\alpha_{2n}}}\left(\frac{d^{a_{2n}}\hat{\rho}_n(\psi)}{d\psi^{a_{2n}}}\right)_{\psi=0}\right]\hspace*{2cm}
\label{dfgqg}
\end{eqnarray}
correspond to a spherically symmetric density of the form 
\begin{equation}
\rho(\psi,r)=\sum\limits_{n=0}^{m}
\tilde{\rho}_n(\psi)r^{2n\beta_{1n}}+ 
\sum\limits_{n=0}^{m}\hat{\rho}_n(\psi)r^{2n\beta_{2n}}/(1+r^2/r_a^2)^{n\beta_{2n}+1} 
\label{rhogqg}
\end{equation}
 with $n\beta_{in}>-1,$ where
$B_{in}=[(2\pi)^{3/2}2^{n\beta_{in}}\Gamma(n\beta_{in}+1)\Gamma(1-\alpha_{in})]^{-1},$ 
$\alpha_{in}=n\beta_{in}-a_{in}+3/2$ and $a_{in}$ is 
a non-negative integer such that $0\leq\alpha_{in}<1$ for $i=1,2$ and $n=0,1,\cdots,m.$

Finally, the velocity dispersions $\sigma_r^2(\psi,r)$ and $\sigma_T^2(\psi,r)$ can be also obtained as  
\begin{equation}
\sigma_r^2(\psi,r)=\frac{1}{\rho(\psi,r)}\sum\limits_{n=0}^{m}
\left[r^{2n\beta_{1n}}\int_0^\psi \tilde{\rho}_n(\psi^\prime)d\psi^\prime+
\frac{r^{2n\beta_{2n}}}{(1+r^2/r_a^2)^{n\beta_{2n}+1}}\int_0^\psi \hat{\rho}_n(\psi^\prime)d\psi^\prime\right] 
\label{vdrqg}
\end{equation} 
and 
\begin{equation}
\sigma_T^2(\psi,r)=\frac{1}{\rho(\psi,r)}\sum\limits_{n=0}^{m}
\left[(n\beta_{1n}+1)r^{2n\beta_{1n}}\int_0^\psi \tilde{\rho}_n(\psi^\prime)d\psi^\prime+
\frac{(n\beta_{2n}+1)r^{2n\beta_{2n}}}{(1+r^2/r_a^2)^{n\beta_{2n}+1}}\int_0^\psi \hat{\rho}_n(\psi^\prime)d\psi^\prime\right]
\label{vdvtqg}
\end{equation} 
for  any DF derived from the spherical density 
given by  (\ref{rhogqg}).  

\section{Application to spherical densities}
\label{appli}
Some anisotropic DFs for spherically symmetric densities are given by using 
formulae obtained in {Sect.}~\ref{anisdf}. Different anisotropic Plummer models 
 are given in {Sect.}~\ref{apm}, together with 
the ratios of their radial and transverse velocity dispersions 
Then anisotropic H\'{e}non isochrone models 
appear in {Sect.}~\ref{ahm} 
and anisotropic $\gamma$-models in {Sect.} \ref{agm}. 
For convenience of calculation, dimensionless quantities are used for all 
the spherical models considered, and it is assumed in this section that 
$Q=\varepsilon-L^2/2$ (or say, take $r_a=1$ in the previous section).

\subsection{Anisotropic Plummer models}
\label{apm}
The well-known Plummer model is given by the dimensionless potential-density pair: 
\begin{equation}
\psi(r)=1/\sqrt{1+r^2}, 
\label{ppsi}
\end{equation}
\begin{equation}
\rho(r)=[3/(4\pi)]/(1+r^2)^{5/2}. \label{prho}
\end{equation}
\begin{figure}[tb]
\psfrag{E}{$\varepsilon$}
\psfrag{L2}{$L^2$}
\psfrag{m3}{$m=3$}
\psfrag{m5}{$m=5$}
\psfrag{m7}{$m=7$}
\centering\includegraphics[bb=16 16 210 221,totalheight=50mm,width=0.4\linewidth]{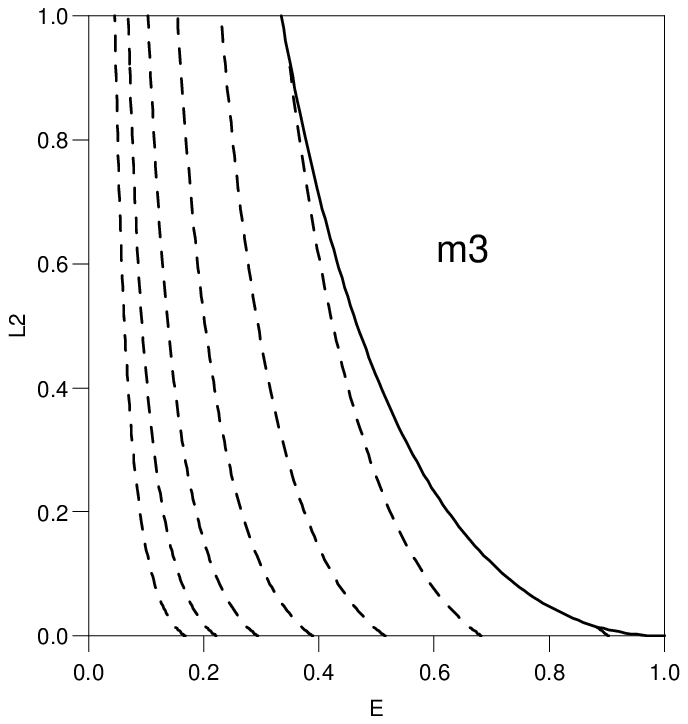}~~
\includegraphics[bb=16 16 210 221,totalheight=50mm,width=0.4\linewidth]{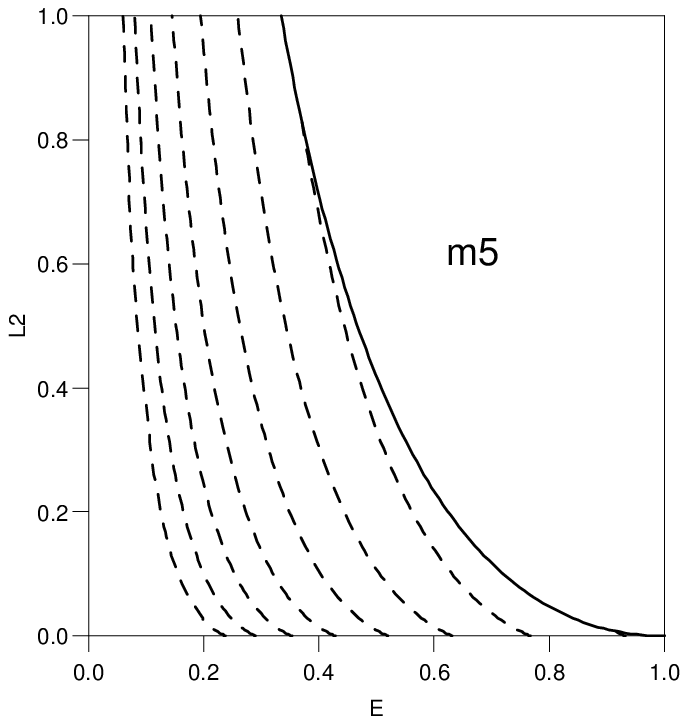}
\includegraphics[bb=16 16 210 221,totalheight=50mm,width=0.4\linewidth]{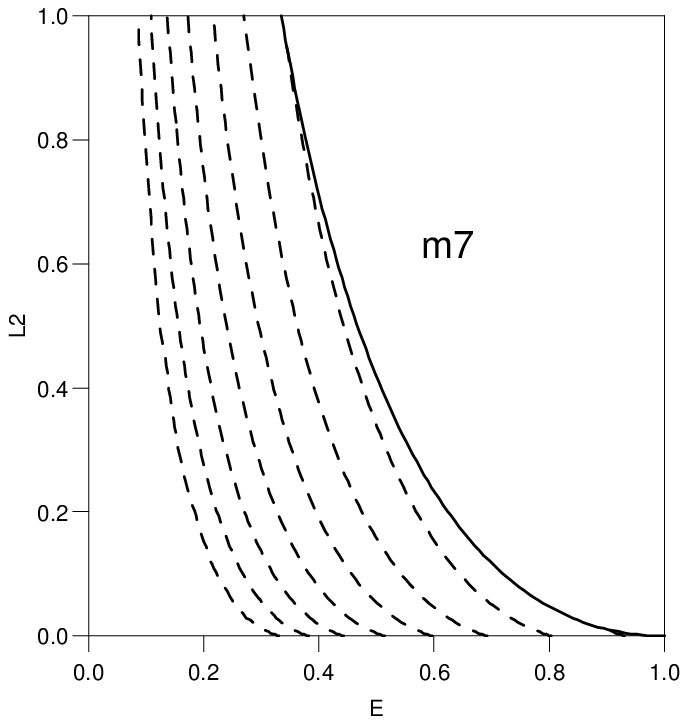}
\caption{The contours of the DFs given by (\ref{pdf}) with  $m=3,5,7.$ 
Here and below, dashed curves are contours and 
the solid curve is the boundary of the physical domain.  
Successive contour levels  differ by factors of $0.08.$ }
\label{pmcon}
\end{figure}

First, DFs of the form given in {Sect.} \ref{anisdf1} are considered for the Plummer model. 
Eq.~(\ref{prho}) can be expressed as  $\rho(\psi,r)=[3/(4\pi)]\psi^{5+2m}(1+r^2)^m,$ 
where $m$ is a positive integer. Then, by (\ref{df0}), 
the anisotropic Plummer model is given by 
\begin{equation}
f(\varepsilon,L)=\frac{3\Gamma(2m+6)\varepsilon^{2m+7/2}}{2(2\pi)^{5/2}}
\left[\sum\limits_{n=0}^{m}\frac{m!}{\Gamma(2m+9/2-n)(m-n)!(n!)^2}
\left(\frac{L^2}{2\varepsilon}\right)^n\right]
\label{pdf}
\end{equation}
for $\varepsilon>0.$  It can be easily proved that equation 
(\ref{pdf}) is in agreement with that given by Dejonghe (1986). 
Figure \ref{pmcon} shows that the contours given by (\ref{pdf}) resemble 
those of the even DFs for some axisymmetric models 
and that the larger the parameter $m,$ the more anisotropic the model. 
Furthermore, by (\ref{vdr}) and (\ref{vdvt}),   
the velocity dispersion ratio $\sigma_r^2(\psi,r)/\sigma_T^2(\psi,r)$ 
for the anisotropic model given by (\ref{pdf}) 
can be obtained as follows:
$\sigma_r^2(\psi,r)/\sigma_T^2(\psi,r)=(1+r^2)/[1+(m+1)r^2].$ 
This is a very good analytical proof of the anisotropic property shown by Figure \ref{pmcon}. 
\begin{figure}[tbh]
\psfrag{E}{$\varepsilon$}
\psfrag{L2}{$L^2$}
\psfrag{c=0.25}{$c=0.25$}
\psfrag{c=0.5}{$c=0.5$}
\centering\includegraphics[bb=16 16 210 221,totalheight=50mm,width=0.4\linewidth]{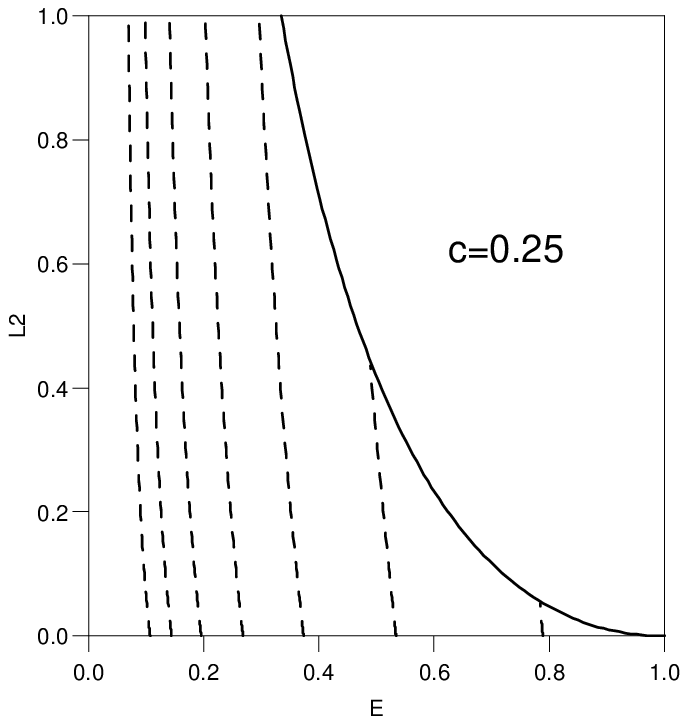}~~
\includegraphics[bb=16 16 210 221,totalheight=50mm,width=0.4\linewidth]{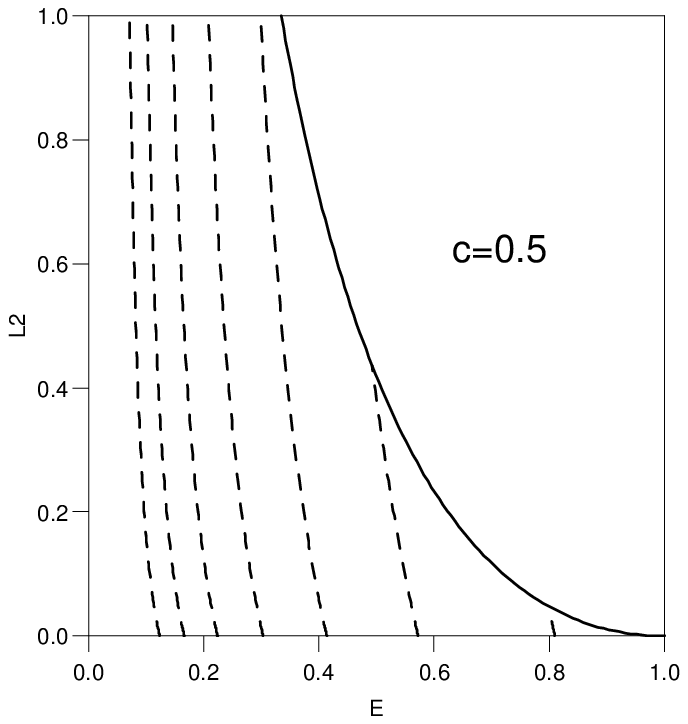}
\caption{The contours of the DFs given by (\ref{pdf0}) with  $c=0.25,0.5.$ Successive contour levels  differ by factors of $0.2.$}
\label{pcfhcon}
\end{figure}

Eq.~(\ref{prho}) can also be rewritten as  $\rho(\psi,r)=[3/(4\pi)]\psi^{7}(1+cr^2)/[\psi^2+c(1-\psi^2)]$ for $c>0,$ 
giving by (\ref{df0}), the anisotropic DF of the Plummer model as
\begin{eqnarray}
f(\varepsilon,L)=\frac{3}{2^{5/2}\pi^3}\int_0^\varepsilon 
\frac{\psi^5\{10(c-1)^2\psi^4+3c[9(1-c)\psi^2+7c]\}}
{[c+(1-c)\psi^2]^3\sqrt{\varepsilon-\psi}}d\psi  \hspace*{3.9cm}\nonumber \\
 +\frac{9cL^2}{2^{7/2}\pi^3}\int_0^\varepsilon 
\frac{\psi^4\{10(1-c)^3\psi^6+39c(1-c)^2\psi^4+7c^2[5c+8(1-c)\psi^2]\}}
{[c+(1-c)\psi^2]^4\sqrt{\varepsilon-\psi}}d\psi \hspace*{0.8cm}
\label{pdf0}
\end{eqnarray}
for $\varepsilon>0.$ Eq.~(\ref{pdf0}) can be calculated analytically 
in terms of generalized hypergeometric functions. However, 
it is not generally easy to calculate numerically 
these hypergeometric functions. To evaluate the DFs, 
it is necessary to estimate the integrals in (\ref{pdf0}). Figure \ref{pcfhcon} 
shows the contours of the two different anisotropic DFs.  
Notice that $c$ is an anisotropic parameter ranging from $0$ to $1.$  
The larger the parameter $c,$ the more anisotropic the model. 
When $c=0,$ it degenates to be an isotropic model; when $c=1,$ 
it becomes the same anisotropic model as given by (\ref{pdf}) with the parameter $m=1.$ 
The velocity dispersion ratio $\sigma_r^2(\psi,r)/\sigma_T^2(\psi,r)$ 
for the anisotropic model defined by (\ref{pdf0}) 
can be further given to be of the simple form  
$\sigma_r^2(\psi,r)/\sigma_T^2(\psi,r)=(1+cr^2)/(1+2cr^2).$
\begin{figure}[tbh]
\psfrag{E}{$\varepsilon$}
\psfrag{L2}{$L^2$}
\psfrag{c=0.25}{$c=0.25$}
\psfrag{c=0.5}{$c=0.5$}
\centering
\includegraphics[bb=16 16 210 221,totalheight=50mm,width=0.4\linewidth]{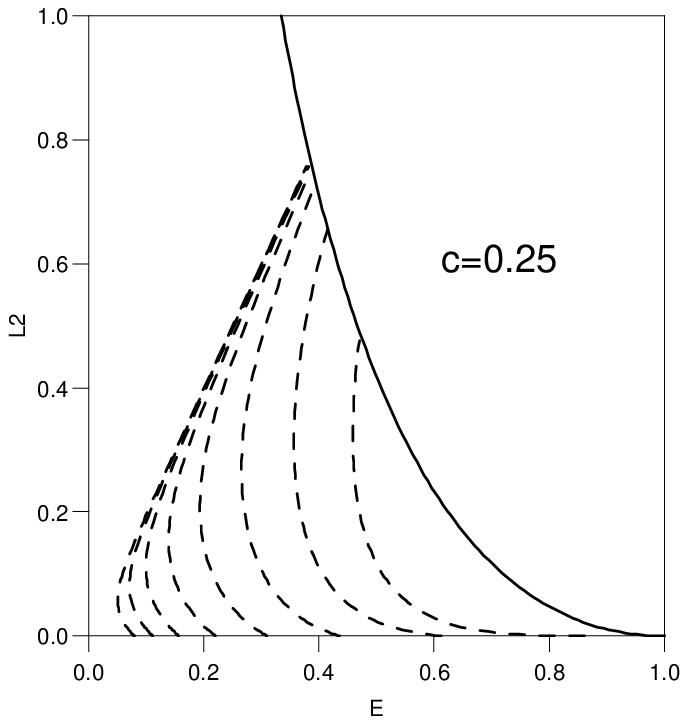}~~
\includegraphics[bb=16 16 210 221,totalheight=50mm,width=0.4\linewidth]{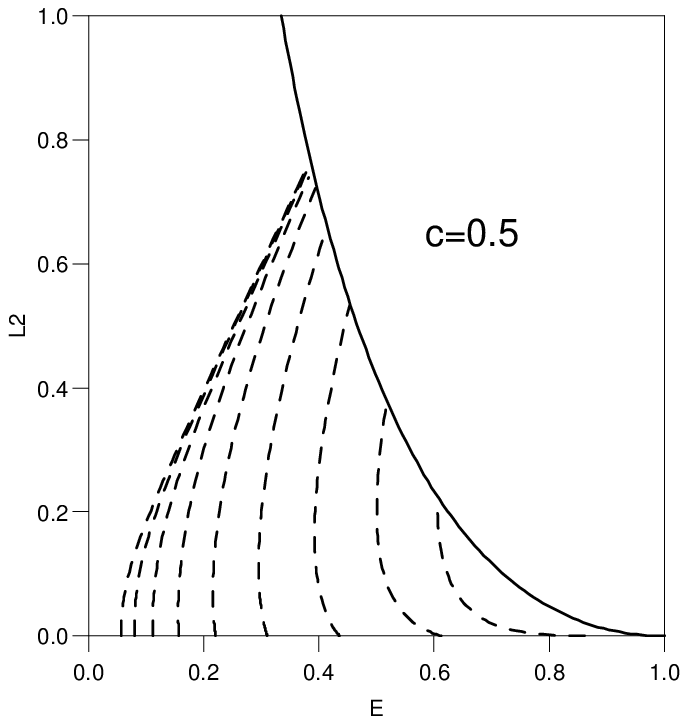}
\caption{The contours of the DFs given by (\ref{pdfq}) with  $c=0.25,0.5.$ Successive contour levels  differ by factors of $0.6.$}
\label{pqcfhcon}
\end{figure}

Next, the DFs of the form given in {Sect.} \ref{anisdf2} are investigated for the Plummer model. 
The density defined by (\ref{prho}) can be expressed as
\begin{equation}
\rho(\psi,r)=[3\psi^3/(4\pi)][c/(1+r^2)+(1-c)(1-\psi^2)^{-1}r^2/(1+r^2)^2]
 \label{prhoq}
\end{equation}
for $0\leq c\leq 1.$ By (\ref{dfq0}), the anisotropic DF for equation (\ref{prhoq}) can be obtained as 
\begin{equation}
f(Q,L)=\frac{3}{2^{7/2}\pi^3} 
\left[8cQ^{3/2}+3(1-c)L^2\int_0^Q\frac{1+6\psi^2+\psi^4}{(1-\psi^2)^4}\frac{d\psi}{\sqrt{Q-\psi}}\right],
\label{pdfq}
\end{equation}
for $Q>0.$ The integral in eq.~(\ref{pdfq}) can 
be calculated analytically 
in terms of hypergeometric functions, and for any $c\in [0,1)$ and $L>0,$ 
the anisotropic DF given by (\ref{pdfq})  
increases like $(1-Q)^{-7/2}$ as $Q\to 1.$  When $c=1,$ (\ref{pdfq})  
is the anisotropic DF of 
Ossipkov-Merritt type (OM).  It was found earlier by Kuzmin and Veltmann (1967a). 
Figure \ref{pqcfhcon} 
displays the contours of the two different DFs obtained 
 by estimating directly the real integral in (\ref{pdfq}). 
Obviously, all the contours plotted here 
lie in the area of $Q>0$ in the physical domain 
since the DFs defined by (\ref{pdfq}) are zero out of this area.    
But, the smaller the anisotropic parameter $c,$ the more anisotropic the model.
By (\ref{vdrq}) and (\ref{vdvtq}), the velocity dispersion ratio $\sigma_r^2(\psi,r)/\sigma_T^2(\psi,r)$ 
for the anisotropic model defined by (\ref{pdfq}) 
can be also given in the following form 
\begin{equation}
\frac{\sigma_r^2(\psi,r)}{\sigma_T^2(\psi,r)}
=\frac{(1+r^2)\{2c\psi^4(1+r^2)-(1-c)r^2[\psi^2+\ln(1-\psi^2)]\}}
{2\{c\psi^4(1+r^2)-(1-c)r^2[\psi^2+\ln(1-\psi^2)]\}}.
 \label{pvdrvtratioq}
\end{equation} 
Inserting (\ref{ppsi}) into  (\ref{pvdrvtratioq}) gives 
 \begin{equation}
\frac{\sigma_r^2(r)}{\sigma_T^2(r)}\equiv\frac{\sigma_r^2(\psi,r)}{\sigma_T^2(\psi,r)}
=\frac{(1+r^2)\{2c-(1-c)r^2[1+\ln\left(\frac{r^2}{1+r^2}\right)]\}}
{2\{c-(1-c)r^2[1+\ln\left(\frac{r^2}{1+r^2}\right)]\}}.
 \label{pvdrvtratioqr}
\end{equation} 
By using the anisotropy parameter defined by Binney (1980), 
it can be also found from (\ref{pvdrvtratioqr}) with $c=1$ that the galaxy 
is isotropic at the centre and becomes increasingly radially 
anisotropic with radius and that the velocity distribution is arbitrarily 
close to one made entirely of radial orbits 
at sufficiently large $r$ (Cuddeford 1991). 

\subsection{Anisotropic H\'{e}non isochrone models}
\label{ahm}
H\'{e}non's (1959a, b) isochrone model has the dimensionless potential-density pair
\begin{equation}
\psi(r)=\frac{1}{1+\sqrt{1+r^2}},
\label{hpsi}
\end{equation}
\begin{equation}
\rho(r)=\frac{1}{4\pi}\frac{3(1+\sqrt{1+r^2})+2r^2}{(1+\sqrt{1+r^2})^3(\sqrt{1+r^2})^3}. 
\label{hrho}
\end{equation} 
Some anisotropic DFs for such model were found by Kuzmin and Veltmann (1967b, 1973). 
Now let us show the other anisotropic DFs for the isochrone model. 
Eq.~(\ref{hrho}) can be expressed as 
\begin{equation}
\rho(\psi,r)=\frac{\psi^5(3+2\psi r^2)}{4\pi(1-\psi)^3},
\label{hrho01}
\end{equation} 
thus, by (\ref{df0}), giving the anisotropic DF as
\begin{equation}
f(\varepsilon,L)=\frac{3}{2^{5/2}\pi^3}\left[\int_0^\varepsilon
\frac{\psi^3(10-5\psi+\psi^2)d\psi}{(1-\psi)^5\sqrt{\varepsilon-\psi}}
+L^2\int_0^\varepsilon\frac{\psi^3(20-15\psi+6\psi^2-\psi^3)d\psi}{(1-\psi)^6\sqrt{\varepsilon-\psi}}
\right],
\label{hdf}
\end{equation}
which can be expressed in terms of elementary functions (See Appendix A). 
Figure \ref{hcon3}(a) shows the contours of the DF given by (\ref{hdf}).  
\begin{figure}[tbh]
\psfrag{E}{$\varepsilon$}
\psfrag{L2}{$L^2$}
\psfrag{hcon}{$(a)$}
\psfrag{h1con}{$(b)$}
\psfrag{h2con}{$(c)$}
\centering
\includegraphics[bb=16 16 210 221,totalheight=50mm,width=0.4\linewidth]{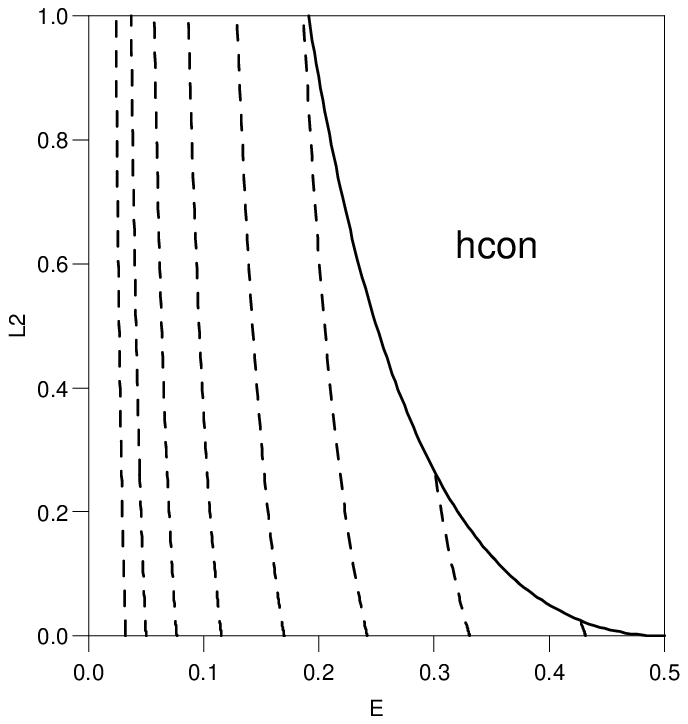}~~
\includegraphics[bb=16 16 210 221,totalheight=50mm,width=0.4\linewidth]{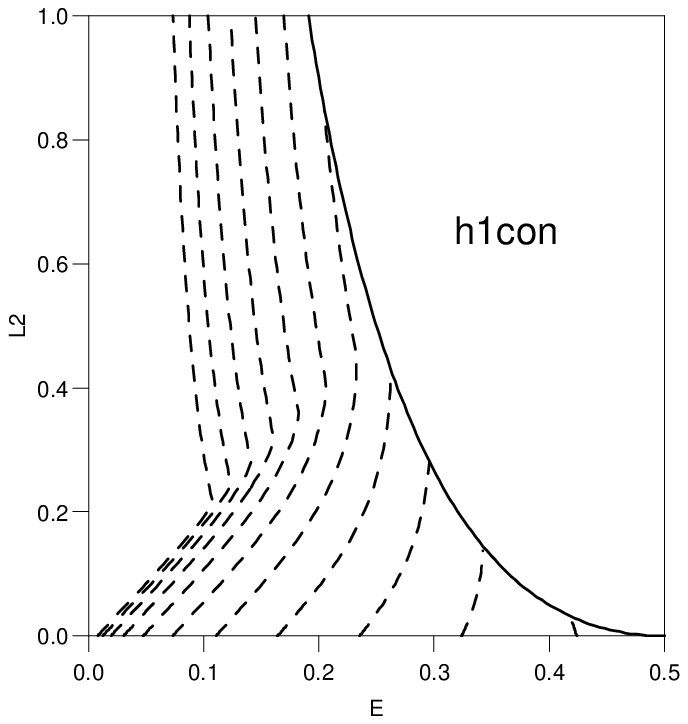}
\includegraphics[bb=16 16 210 221,totalheight=50mm,width=0.4\linewidth]{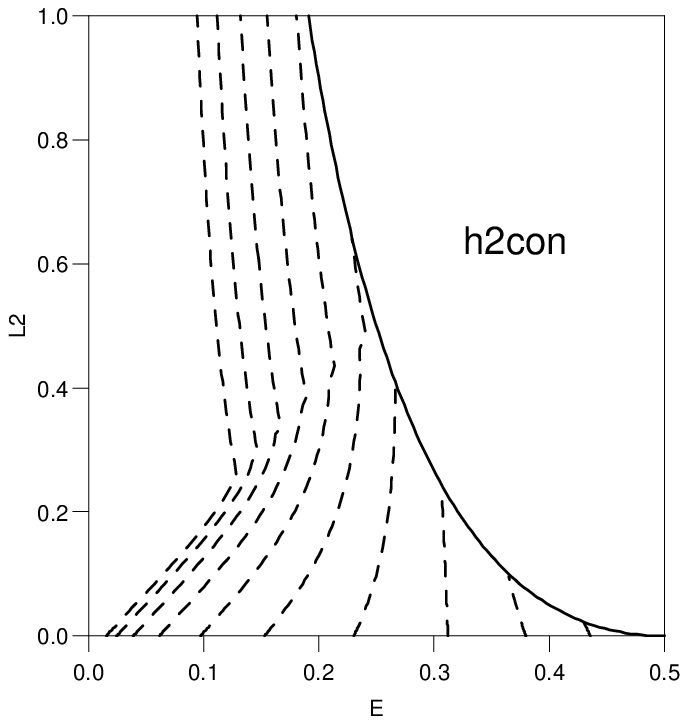}
\caption{The contours of the DFs for H\'{e}non's model. (a), (b) and (c) 
are for (\ref{hdf}), (\ref{hdfg}) and (\ref{hdfg0}),  respectively. 
In (a), successive contour levels  differ by factors of $0.2.$ 
In (b) and (c),  successive contour levels  differ by factors of $0.5.$}
\label{hcon3}
\end{figure}

Eq.~(\ref{hrho}) can be expressed as 
\begin{equation}
\rho(\psi,r)=\frac{1}{4\pi}\left[\frac{3\psi^3}{1-\psi}\frac{1}{1+r^2}
+\frac{2\psi^6r^2}{(1-\psi)^3}\right],
\label{hrhog}
\end{equation}   
and
then it follows from  (\ref{dfgqg}) that 
the anisotropic DF can be obtained as
\begin{equation}
f(\varepsilon,Q,L)=\frac{3}{2^{5/2}\pi^3}
\left[\int_0^{\hat{Q}}\frac{\psi(3-3\psi+\psi^2)d\psi}{(1-\psi)^3\sqrt{\hat{Q}-\psi}}
 +L^2\int_0^\varepsilon\frac{\psi^3(20-15\psi+6\psi^2-\psi^3)d\psi}{(1-\psi)^6\sqrt{\varepsilon-\psi}}
\right],
\label{hdfg}
\end{equation}
where $\hat{Q}=\max(Q,0).$ 
 The integral in (\ref{hdfg}) can be calculated analytically in terms of elementary functions (See Appendix A). 
Figure~\ref{hcon3}(b) displays the contours of 
the DF given by (\ref{hdfg}). 

The density (\ref{hrho}) can be also expressed as 
\begin{equation}
\rho(\psi,r)=\frac{1}{4\pi}\left[\frac{3\psi^6}{(1-\psi)^3}+\frac{3\psi^3}{1+r^2}
+\frac{2\psi^6r^2}{(1-\psi)^3}\right],
\label{hrhogq}
\end{equation}   
and, corresponding to it, by (\ref{dfgqg}), the anisotropic DF  is 
\begin{eqnarray}
f(\varepsilon,Q,L)=\frac{3}{2^{5/2}\pi^3}\left[3\int_0^\varepsilon
 \frac{\psi^4(5-4\psi+\psi^2)d\psi}{(1-\psi)^5\sqrt{\varepsilon-\psi}}
 +4\hat{Q}^{3/2}\right. \hspace*{5.7cm}\nonumber\\
\left.
+L^2\int_0^\varepsilon\frac{\psi^3(20-15\psi+6\psi^2-\psi^3)d\psi}{(1-\psi)^6\sqrt{\varepsilon-\psi}} 
\right],\hspace*{5cm}
\label{hdfg0}
\end{eqnarray}
where $\hat{Q}=\max(Q,0).$ 
The two integrals at the right side of equation (\ref{hdfg0}) 
can be also expressed in terms of elementary functions (See Appendix A). 
Figure \ref{hcon3}(c) illustrates the contours of the DF given  
by (\ref{hdfg0}).  

\subsection{Anisotropic $\gamma$-models}
\label{agm}
The dimensionless potential-density pair of the $\gamma$-model (Kuzmin, Veltmann, Tenjes 1986;  
Dehnen 1993; Saha 1993; Tremaine et al.~1994) is
\begin{equation}
\psi(r)=\left\{\begin{array}{ll} [1-r^{2-\gamma}/(r+1)^{2-\gamma}]/(2-\gamma), & \gamma\not=2 \cr
\ln[(r+1)/r], & \gamma=2 \end{array}\right.
\label{gpsi}
\end{equation}
\begin{equation}
\rho(r)=\frac{3-\gamma}{4\pi}\frac{1}{r^\gamma(r+1)^{4-\gamma}}. 
\label{grho}
\end{equation}

In the same way as for the Plummer model, by (\ref{df0}), it can be shown
 that the anisotropic DFs of the form of Sect.~\ref{anisdf1} 
can be obtained 
if the density defined by (\ref{grho}) is expressed as
\begin{equation}
\rho(\psi,r)=\frac{3-\gamma}{4\pi}\frac{(1-y)^{4+2m}(1+r^2)^m}
{ y^\gamma[(1-y)^2+y^2]^m},
\label{grhor}
\end{equation}
where $m$ is a positive integer and $y$ is written as 
\begin{equation}
y=\left\{\begin{array}{ll} [1-(2-\gamma)\psi]^{1/(2-\gamma)}, & \gamma\not=2 \\
 e^{-\psi}, & \gamma=2 \end{array}\right.  .
\label{y}
\end{equation}
In fact, it is known from (\ref{gpsi}) that $y=r/(r+1).$ 
These anisotropic DFs cannot be expressed in terms of  
 elementary functions or generalized hypergeometric functions 
although they can be calculated numerically. 
\begin{figure}[ht]
\psfrag{E}{$\varepsilon$}
\psfrag{L2}{$L^2$}
\centering
\includegraphics[bb=16 16 210 221,totalheight=50mm,width=0.7\linewidth]{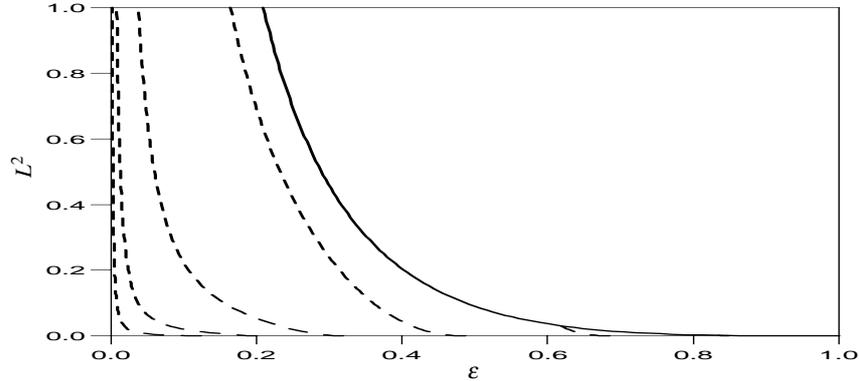}
\caption{The contours of the anisotropic Hernquist DFs (\ref{gdf}). 
Successive contour levels  differ by factors of $0.1.$}
\label{hernquist}
\end{figure}

Another expression of the density for the $\gamma$-model is 
\begin{equation}
\rho(\psi,r)=\frac{3-\gamma}{4\pi}\frac{(1-y)^{4+2m}(1+r)^{2m}}
{ y^\gamma},
\label{grhod}
\end{equation}
where $y$ is given in (\ref{y}). It can be shown from (\ref{dfg}) that 
the anisotropic DFs corresponding to (\ref{grhod}) can be expressed 
generally in terms of  
 elementary functions or generalized hypergeometric functions. 
In particular, when $\gamma=1,$ it is a model obtained by Kuzmin and Veltmann (1973) and Hernquist (1990) 
and its density can be expressed as      
\begin{equation}
\rho(\psi,r)=\frac{1}{2\pi}\frac{\psi^{4+2m}}{1-\psi}(1+r)^{2m}.
\label{grhoh}
\end{equation}
For example, by use of (\ref{dfg}), a DF of the mass density 
given by (\ref{grhoh}) with $m=1$ is
\begin{equation}
f(\varepsilon,L)=\frac{1}{(2\pi^2)^{3/2}}\left[\int_0^\varepsilon
\frac{h(\psi)d\psi}{\sqrt{\varepsilon-\psi}}
+2\sqrt{2}Lh(\varepsilon)+\frac{3L^2}{2}\int_0^\varepsilon
\frac{\psi^3(20-45\psi+36\psi^2+10\psi^3)d\psi}
{(1-\psi)^4\sqrt{\varepsilon-\psi}}\right]
\label{gdf}
\end{equation}
for $\varepsilon>0,$ where $h(\psi)=\psi^4(15-24\psi+10\psi^2)/(1-\psi)^3.$ 
The two integrals in eq.~(\ref{gdf}) can be expressed in terms of  
 elementary functions (See (\ref{id05int}) and (\ref{id06int}) in Appendix A).
The contours of the DF given by (\ref{gdf}) are plotted 
in Figure~\ref{hernquist}. 

By (\ref{dfq0}), the anisotropic DFs of the form in {Sect.} \ref{anisdf2} 
can also be obtained in terms of elementary functions or 
generalized hypergeometric functions in the same way as  
the DFs of Ossipkov-Merritt type are given by Dehnen (1993). 
The detail derivation of these expressions is omitted here. 

By the way,  Baes and Dejonghe (2005) recently 
used the spherical $\gamma$-models 
to investigate the dynamical structure of isotropic spherical galaxies 
with a central black hole. Such work is still very significant.

\section{Conclusions}
\label{con}
Although real galaxies are hardly spherically symmetric,  
it is  a very necessary and significant step to study the self-consistent anisotropic 
DFs for stellar systems with the known spherically symmetric density.  
This is not only because some important properties 
(e.g. surface density and cumulative mass) of 
the spherical model are similar to those of 
the models that can be generated from the spherically symmetric density 
by replacing the 
spherical radius $r$ by an axisymmetric or triaxial radius 
$m=\sqrt{x^2+(y/q_2)^2+(z/q_3)^2},$  but also because both spherical and non-spherical 
models have the same typical behaviours of the dynamical quantities 
in the limits of small  and large radii. Such models are elliposoidal. 
Other classes of flattened models can be constructed by using the equipotential method 
(e.g., Kutusov and Ossipkov 1980; Jiang 2000).

Anisotropic DFs can be obtained for  stellar systems
with known spherically symmetric density as a sum of products 
of functions only of the potential and 
a  special function (or power) only of the radial coordinate, 
i.e. these DFs are a sum of products of functions 
only of a special variable (or the energy)  
and a power only of the 
magnitude of the angular momentum. 
This comes from a combination of the ideas of 
 Eddington and Fricke. 
It is an extension of Eddington's classical solution for 
the isotropic DF of a known spherical density. 
Like his method, it requires that the density be expressed as 
a special function of the potential, and now also of the radial coordinate.  
Also, part of it is a sum of 
real integrals of functions only of the potential. 
Most of these integrals can be calculated analytically in terms of  
 elementary functions or 
generalized hypergeometric functions 
(e.g. Kuzmin and Veltmann 1967b). 

These formulae concerning anisotropic DFs are obtained by use of the Abel 
integral equation, and they are suitable for all such densities that 
allow the real integrals of the potential to be valid. 
The anisotropic DFs of Ossipkov-Merritt type are their examples in special cases. 
After the application to Plummer's spherical model, it is found that 
the anisotropic DFs given by (\ref{pdf}) are in fact the same as 
those given by Dejonghe (1986). Three different anisotropic DFs are given
for H\'{e}non's (1959a, b) isochrone model and 
they can be expressed in terms of elementary functions. 
Many anisotropic DFs of $\gamma$-models,  
expressed in elementary functions or generalized hypergeometric functions,
can also be obtained. 
One can further give formulae of the velocity dispersions 
 for these anisotropic DFs.

These expressions can be extended further to axisymmetric systems. 
It is straightforward to find their analogues for axisymmetric systems  
and they can be used to obtain the even DF of Binney's (BT) logarithmic 
potential although Evans (1993) derived it using Lynden-Bell's (1962) method. 
For the well-known Lynden-Bell (1962) model,  
these analogues degenerate into the method of Fricke (1952). 
 
\begin{acknowledgement}
The first author was supported in part by NSFC 10271121 and by 
SRF for ROCS, SEM. 
The numerical calculations were performed by use of the 
computer cluster at the Department of Mathematics 
of the University of Manchester. 
The first author would like to thank Dr.~David Moss for his 
helpful comments on this paper.  
The second author was supported by Leading Scientific School grant 1078.2003.02. 
The cooperation of authors was supported by joint grants 
of NSFC 10511120278/10611120371 and RFBR 04-02-39026. 
The two authors are very grateful to Professor Konstantin Kholshevnikov 
for his valuable discussions on this work.  
\end{acknowledgement}

\section*{Appendix A:  Explicit expressions of integrals for the anisotropic DFs of both H\'{e}non's model and Hernquist's model} 
\label{appc}
Explicit expressions of integrals for the DFs of the H\'{e}non model 
are given below.
Using the identity
\begin{equation}
\frac{\psi^3(10-5\psi+\psi^2)}{(1-\psi)^5}
=\frac{6}{(1-\psi)^5}-\frac{15}{(1-\psi)^4}+\frac{10}{(1-\psi)^3}-1,
\label{id01}
\end{equation}
the expression  
\begin{eqnarray}
\int_0^\varepsilon \frac{\psi^3(10-5\psi+\psi^2)d\psi}
{(1-\psi)^5\sqrt{\varepsilon-\psi}}
=\frac{(19-106\varepsilon+240\varepsilon^2-48\varepsilon^3)\sqrt{\varepsilon}}
{32(1-\varepsilon)^4} \hspace*{5.5cm}\nonumber\\
+\frac{15(3-12\varepsilon+16\varepsilon^2)\arcsin(\sqrt{\varepsilon})}
{32(1-\varepsilon)^{9/2}}-2\sqrt{\varepsilon} \hspace*{3cm}
\label{id01int}
\end{eqnarray}
can be derived.
Similarly, the identity
\begin{equation}
\frac{\psi^3(20-15\psi+6\psi^2-\psi^3)}{(1-\psi)^6}
=\frac{10}{(1-\psi)^6}-\frac{24}{(1-\psi)^5}+\frac{15}{(1-\psi)^4}-1,
\label{id02}
\end{equation}
gives              
\begin{eqnarray}
\int_0^\varepsilon \frac{\psi^3(20-15\psi+6\psi^2-\psi^3)d\psi}
{(1-\psi)^6\sqrt{\varepsilon-\psi}}
=\frac{(53-330\varepsilon+880\varepsilon^2-352\varepsilon^3
+64\varepsilon^4)\sqrt{\varepsilon}}
{64(1-\varepsilon)^5} \hspace*{2.8cm}\nonumber\\
+\frac{15(5-24\varepsilon+40\varepsilon^2)\arcsin(\sqrt{\varepsilon})}
{64(1-\varepsilon)^{11/2}}-2\sqrt{\varepsilon}. \hspace*{2cm}
\label{id02int}
\end{eqnarray}
Note that $\frac{\psi(3-3\psi+\psi^2)}{(1-\psi)^3}
=\frac{1}{(1-\psi)^3}-1.$ Then 
\begin{equation}
\int_0^Q \frac{\psi(3-3\psi+\psi^2)d\psi}
{(1-\psi)^3\sqrt{Q-\psi}}
=\frac{(5-2Q)\sqrt{Q}}{4(1-Q)^2}
+\frac{3\arcsin(\sqrt{Q})}{4(1-Q)^{5/2}}
-2\sqrt{Q}.
\label{id03int}
\end{equation}
Using the identity
\begin{equation}
\frac{\psi^4(5-4\psi+\psi^2)}{(1-\psi)^5}
=\frac{2}{(1-\psi)^5}-\frac{6}{(1-\psi)^4}+\frac{5}{(1-\psi)^3}
-2+(1-\psi),
\label{id04}
\end{equation}
the result     
\begin{eqnarray}
\int_0^\varepsilon \frac{\psi^4(5-4\psi+\psi^2)d\psi}
{(1-\psi)^5\sqrt{\varepsilon-\psi}}
=\frac{(87-350\varepsilon+464\varepsilon^2-96\varepsilon^3)\sqrt{\varepsilon}}
{96(1-\varepsilon)^4} \hspace*{5.5cm} \nonumber \\
+\frac{5(7-24\varepsilon+24\varepsilon^2)\arcsin(\sqrt{\varepsilon})}
{32(1-\varepsilon)^{9/2}}
-\frac{2\sqrt{\varepsilon}(3+2\varepsilon)}{3} \hspace*{3cm}
\label{id04int}
\end{eqnarray}
is obtained.

Similarly, integrals for the anisotropic DFs of the Hernquist model
can be also expressed as
\begin{eqnarray}
\int_0^\varepsilon\frac{\psi^4(15-24\psi+10\psi^2)d\psi}
{(1-\psi)^3\sqrt{\varepsilon-\psi}}=-
\frac{2}{35}\sqrt{\varepsilon}
(35+70\varepsilon+112\varepsilon^2+160\varepsilon^3)
\hspace*{4.2cm} \nonumber \\ +
\frac{(5-2\varepsilon)\sqrt{\varepsilon}}{4(1-\varepsilon)^2}
+\frac{3\arcsin(\sqrt{\varepsilon})}{4(1-\varepsilon)^{5/2}}
\hspace*{4cm}
\label{id05int}
\end{eqnarray}
and 
\begin{eqnarray}
\int_0^\varepsilon\frac{\psi^3(20-45\psi+36\psi^2-10\psi^3)d\psi}
{(1-\psi)^4\sqrt{\varepsilon-\psi}}=-
\frac{2}{3}\sqrt{\varepsilon}
(3+8\varepsilon+116\varepsilon^2)\hspace*{4.8cm} \nonumber \\+
\frac{(33-26\varepsilon+8\varepsilon^2)\sqrt{\varepsilon}}{24(1-\varepsilon)^3}
+\frac{15\arcsin(\sqrt{\varepsilon})}{24(1-\varepsilon)^{7/2}}.
\hspace*{2cm}
\label{id06int}
\end{eqnarray}


\vspace*{0.8cm}
{\noindent\bf References}\vspace*{0.1cm}
{\small 
\begin{description}
\item[Baes M., Dejonghe H.:] The dynamical structure of isotropic spherical galaxies with a central black hole. 
   Astronomy \& Astrophysics, {\bf 432}, 411-422 (2005). 
\item[Binney J.:] The radius-dependence of velocity dispersion in elliptical galaxies. 
Monthly Notices of the Royal Astronomical Society, {\bf 190}, 873-880  (1980). 
\item[Binney J., Tremaine S.:] Galactic Dynamics. Princeton Univ., Princeton (1987). 
\item[Bouvier P.:] Sur le structure des amas globulaires.
     Archives des Sciences (Gen{\'e}ve), {\bf 15}, 163-173 (1962).
\item[Bouvier P.:] Distribution des vitesses dans un syst{\`e}me sph{\'e}rique   
     quau permanent.
     Archives des Sciences (Gen{\'e}ve), {\bf 16}, 195-210 (1963).
\item[Camm G.~L.:] Self-gravitating star systems. II. 
Monthly Notices of the Royal Astronomical Society, {\bf 112}, 155-176 (1952). 
\item[Cuddeford P.:]  An analytic inversion for anisotropic spherical galaxies. 
Monthly Notices of the Royal Astronomical Society, {\bf 253}, 414-426 (1991). 
\item[Dehnen W.:] A family of potential-density pairs for spherical galaxies and bulges. 
Monthly Notices of the Royal Astronomical Society, {\bf 265}, 250-256 (1993).
\item[Dejonghe H.:] Stellar dynamics and the description of stellar systems. 
Physics Reports, {\bf 133}, 217-313 (1986). 
\item[Dejonghe H.:] A completely analytical family of anisotropic Plummer. 
Monthly Notices of the Royal Astronomical Society, {\bf 224}, 13-39  (1987). 
\item[Dejonghe H., Merritt D.:]  Radial and nonradial stability of spherical stellar systems. 
Astrophysical Journal, {\bf 328}, 93-102  (1988).
\item[Eddington A.~S.:] The distribution of stars in globular clusters. 
Monthly Notices of the Royal Astronomical Society, {\bf 76}, 572-585 (1916). 
\item[Evans N.~W:] Simple galaxy models with massive haloes. 
Monthly Notices of the Royal Astronomical Society, {\bf 260}, 191-201 (1993). 
\item[Fricke W.:] Dynamische Begr\"{u}ndung der Geschwindigkeitsverteilung im Sternsystem. 
Astron.~Nachr., {\bf 280}, 193-216 (1952).
\item[H\'{e}non M.:]  L'amas isochrone: I. 
Annales d'Astrophysique, {\bf 22}, 126-139 (1959a). 
\item[H\'{e}non M.:]  II. Le calcul des orbites. 
Annales d'Astrophysique, {\bf 22}, 491-498 (1959b). 
\item[Hernquist L.:] An analytical model for spherical galaxies and bulges.  
Astrophysical Journal, {\bf 356}, 359-364 (1990). 
\item[Hunter C.:] Determination of the distribution function of an elliptical galaxy. 
Astronomical Journal, {\bf 80}, 783-793 (1975). 
\item[Hunter C., Qian E.:] Two-integral distribution functions for axisymmetric galaxies. 
Monthly Notices of the Royal Astronomical Society, {\bf 262}, 401-428 (1993). 
\item[Jiang Z.:] Flattened Jaffe models for galaxies. 
Monthly Notices of the Royal Astronomical Society, {\bf 319}, 1067-1078 (2000). 
\item[Kalnajs A.~J.:] Dynamics of Flat Galaxies. III. Equilibrium Models. Astrophysical Journal, {\bf 205}, 751-761 (1976).
\item[Kent S.~M., Gunn J.~E.:] The dynamics of rich clusters of galaxies. I - The Coma cluster. 
Astronomical Journal,  {\bf 87}, 945-971 (1982). 
\item[Kutuzov S.~A., Ossipkov L.~P.:] A generalized model for the three-dimensional gravitational potential of stellar systems. 
Pis'ma v Astronomicheskij Zhurnal, {\bf 57}, 28-37 (1980) (English translations: 
Soviet Astronomy Letters, {\bf 24}, 17-22, (1981)). 
\item[Kuzmin G.~G., Veltmann \"{U}.-I.K.:] Hydrodynamic models of spherical stellar systems. 
 W.~Struve Tartu Astrof\"u\"us.~Obs.~Publ., {\bf 36}, 3-47 (1967a). 
\item[Kuzmin G.~G., Veltmann \"{U}.-I.K.:] Lindblad diagram and isochronic
      models. W.~Struve Tartu Astrof\"u\"us.~Obs.~Publ., {\bf 36}, 470-507 (1967b).
\item[Kuzmin G.~G., Veltmann \"{U}.-I.K.:] Generalized isochrone models for  
      spherical stellar systems.
      Dynamics of Galaxies and Star Clusters, Nauka, Alma-Ata, 82-87 (1973)  
      (English translations: Galactic Bulges (IAU Symp. 153), Kluwer, Dordrecht,  
       363-366 (1993)).
\item[Kuzmin G.~G., Veltmann \"{U}.-I.K., Tenjes P.~L.:]  Quasi-isothermal 
      models of spherical stellar systems. Application to the galaxies M~87 and  
      M~105. 
      W.~Struve Tartu Astrof\"{u}\"{u}s.~Obs. Publ., {\bf 51}, 232-242 (1986). 
\item[Louis P.~D.:] Models for spherical stellar systems with isotropic cores and anisotropic haloes. 
Monthly Notices of the Royal Astronomical Society, {\bf 261}, 283-298 (1993). 
\item[Lynden-Bell D.:] Stellar dynamics: Exact solution of the self-gravitation equation. 
Monthly Notices of the Royal Astronomical Society, {\bf 123}, 447-458 (1962). 
\item[Merritt D.:] Spherical stellar systems with spheroidal velocity distributions. 
Astronomical Journal, {\bf 90}, 1027-1037 (1985).
\item[Ogorodnikov K.~F.:] Dynamics of Stellar Systems, Pergamon Press, London 
      (1965).
\item[Ossipkov L.~P.:] Some problems of the theory of self-consistent models for
       star clusters.
       Star Clusters, Urals Univ.~Press, Sverdlovsk, 72-89 (1979a).
\item[Ossipkov L.~P.:] Spherical systems of gravitating bodies with an  
      ellipsoidal velocity distribution. 
      Pis'ma v Astronomicheskij Zhurnal, {\bf 5}, 77-80 (1979b)
      (English translations: Soviet Astronomy Letters, {\bf 5}, 42-44).
\item[Plummer H.~C.:] On the problem of distribution in globular star clusters. 
Monthly Notices of the Royal Astronomical Society, {\bf 71}, 460-470 (1911). 
\item[Qian E., Hunter C.:] Anisotropic distribution functions for spherical 
      galaxies.
      {Astron.~}{Astrophys.~}Transact., {\bf 7}, 201-206 (1995). 
\item[Saha P.:] Designer basis functions for potentials in galactic dynamics. 
Monthly Notices of the Royal Astronomical Society, {\bf 262}, 1062-1064 (1993). 
\item[Shiveshwarkar S.~W.:] Remarks on some theorems in the dynamics of a steady stellar system. 
Monthly Notices of the Royal Astronomical Society, {\bf 96}, 749-757 (1936). 
\item[Tremaine S.~et al.:] A family of models for spherical stellar systems. 
Astronomical Journal, {\bf 107}, 634-644 (1994). 
\item[Veltmann \"{U}.-I.K.:] Constructing models for spherical star systems with 
      given space density. 
      Tartu {Astron.~}{Obs.~}Publ., {\bf 33}, 387-415 (1961). 
\item[Veltmann \"{U}.-I.K.:] On phase density of spherical stellar systems.
      {W.~}Struve Tartu {Astrof\"{u}\"{u}s.~}{Obs.~}Publ., {\bf 35}, 5-26 (1965). 
\item[Veltmann \"{U}.-I.K.:] Phase space  models for star clusters.
       Star Clusters, Urals {Univ.~}Press, Sverdlovsk,  50-71 (1979).
\item[Veltmann \"{U}.-I.K.:] Gravitational potential, space density and phase  
      density of star clusters. 
      W.~Struve Tartu Astrof\"{u}\"{u}s.~Obs. Publ., {\bf 48}, 232-261 (1981). 
\end{description}
}
\end{document}